# In-situ formation of SiGe alloy by electron beam evaporation and the effect of post deposition annealing on the energy band gap

Twisha Tah,[a] Ch. Kishan Singh[*,a] S. Amirthapandian,[a] K. K. Madapu,[a] A. Sagdeo,[b,c] S. Ilango,[a] T. Mathews[a] and S. Dash[d]

[a] *Material Science Group, Indira Gandhi Centre for Atomic Research, HBNI, Kalpakkam – 603102, India*

[b] *Synchrotrons Utilization Section, Raja Ramanna Centre for Advanced Technology, Indore-452013, India*

[c] *Homi Bhabha National Institute, Anushakti Nagar, Mumbai 400094, India*

[d] *CBCMT, VIT University, Vellore – 632014, India*

## Abstract

We report the synthesis of polycrystalline (*poly*)-SiGe alloy thin films through solid state reaction of Si/Ge multilayer thin films on Si and glass substrates at low temperature of 500 °C. The pristine thin film was deposited using electron beam evaporation with optimized in-situ substrate heating. Our results show the co-existence of amorphous Si (*a*-Si) phase along with the *poly*-SiGe phase in the pristine thin film. The *a*-Si phase was found to subsume into the SiGe phase upon post deposition annealing in the temperature range from 600 to 800 °C. Additionally, dual energy band gaps could be observed in the optical properties of the annealed poly-SiGe thin films. The stoichiometric evolution of the pristine thin film and its subsequent effect on the band gap upon annealing are discussed on the basis of diffusion characteristics of Si in *poly*-SiGe.

**Keywords**      : Poly-SiGe thin film; In-situ crystallization; band gap; diffusion

**PACS**          : *81.05.Bx; 81.15.Dj; 81.05.Hd; 68.35.Fx*

---

[*]Email: kisn@igcar.gov.in; kisnsingh@gmail.com and twishatah1991@gmail.com

## 1. Introduction

For the past several decades, the prime material for absorber layer of solar cells in photovoltaic (PV) industries has been elemental Si, both in its amorphous (*a*-Si) as well as crystalline (*c*-Si) form [1,2]. Si is also the basic material in high performance optoelectronic and microelectronic devices like photodiode, active matrix-LCD, image sensors, integrated electronics [3]. It is largely due to abundance and cost effectiveness in large scale production of Si. However, Si has inherently low absorption co-efficient in the Infra Red (IR) region of the solar spectrum and devices based on Si have low efficiency and light induced degradation due to Staebler-Wronski-Effect [4,5]. Due to these shortcomings, SiGe has been investigated as an alternative to Si for the above applications for past few decades [6]. The attractive properties of SiGe results from the addition of electrically and optically superior Ge that makes the lattice constant and energy band gap tunable in a broad range. SiGe as a result have high absorption co-efficient and its optical response can be improved in the IR region. Hence, particularly for PV industries, the use of SiGe increases the effective absorption of the solar spectrum and increases the efficiency of the thin film solar cell. The use of amorphous SiGe (*a*-SiGe) as absorber material in solar cell is reported to result in the increase of long wavelength response in Quantum Efficiency (QE) up to 950 nm [7]. The longer wavelength response in QE can be further extended beyond the value reported with *a*-SiGe by using microcrystalline SiGe (*mc*-SiGe) instead of *a*-SiGe [8]. Apart from crystallinity, the spectral response depends on the Ge content in the SiGe alloy. This was demonstrated when J. Ni *et al.* [9] reported achieving an spectral response of up to 1300 nm using *mc*-SiGe with a Ge fraction of ≈ 77 %. Hence, good quality Ge rich *c*-SiGe is desirable for PV applications. Further, the melting temperature $T_m$ of SiGe is lower than Si as it lies between $T_{m(Ge)}$ = 940 °C and $T_{m(Si)}$ = 1414 °C [10]. As such, deposition, crystallization, grain growth and dopant activation in SiGe occur at relatively lower temperature compared to Si and find use in applications where lower thermal budget is desirable. Herein, the synthesis of Ge rich polycrystalline (*poly*)-SiGe at low temperature is challenging. The synthesis methods generally used to achieve *poly*-Si, Ge and SiGe are Solid Phase Crystallization (SPC), laser annealing, electron beam (e-beam) crystallization, Metal Induced Crystallization (MIC), etc [11–15]. Among these techniques, MIC can enable growth of SiGe at the lowest temperature in the range ~ 150 to 400 °C [16,17]. However, there is a chance of degradation in semiconducting properties due to metal contamination in MIC process [18]. In contrast, SPC is the simplest and most cost effective method. It does not involve any toxic metals or gas. Using SPC, *a*-SiGe is reported to transform into *poly*-SiGe at 625 °C after annealing for a period of more than 24 hrs [19]. Earlier, we have reported formation of nanocrystalline (*nc*)-SiGe by ex-situ annealing of Ge/Si (111) thin films by e-beam evaporation [20]. No *nc*-SiGe phase could be observed with in-situ substrate heating at 450 °C during deposition. It could be achieved only upon ex-situ annealing at temperature ≥ 600 °C.

In this paper, we report the synthesis of Ge-rich $Si_xGe_{1-x}$ alloy thin film on Si and Corning 0215 glass substrates using e-beam evaporation method at low temperature of 500 °C. With in-situ substrate heating at 500 °C, crystalline $Si_xGe_{1-x}$ phase could be achieved during the sequential evaporation of Si/Ge multilayer thin film on Si and Corning glass substrates. The stoichiometric evolution of the $Si_xGe_{1-x}$ phase along with the optical characteristics upon post deposition isochronal annealing at higher temperatures ranging from 600 to 800 °C is discussed.

2. **Experimental**

For the present study, thin film samples were prepared by sequential deposition of Ge/Si of 15 nm each with two periods onto cleaned Si (100) and corning glass substrates using e-beam evaporation. Si was chosen as the top layer of the period to prevent loss of Ge due desorption at high temperatures [21,22]. The base pressure in the deposition chamber and the working pressure during deposition were ~ $10^{-8}$ mbar and ~ $10^{-7}$ mbar, respectively. The temperatures of the substrates, $T_s$ were maintained at 500 °C by in-situ heating of the substrate stage and the stage was rotated for film uniformity during the deposition. In the forthcoming discussion, for brevity, the thin film deposited at 500 °C will be hereafter referred to as 'pristine sample'. Post deposition isochronal annealing of the pristine thin films was performed at annealing temperature $T_a$, ranging from 600 to 800 ºC for 5 hrs in $N_2$ ambient. The structural characteristics of all the thin film samples in the present study were investigated using grazing incidence x-ray diffraction (GIXRD) and Raman spectroscopy. The GIXRD measurements were performed with 15.501 keV energy x-rays at 0.3 degree angle of incidence on angle dispersive x-ray diffraction beam line of the Indus-2 Synchrotron facility at RRCAT, Indore. The Raman spectral measurements of all samples in the present study were performed in a Raman microscope (inVia, Renishaw, UK) using green light laser 514.5 nm as the excitation source along with 1800 gr/mm grating and CCD detector. The detailed structural phase information of the thin film sample was further investigated with high resolution transmission electron microscopy (HRTEM) imaging and selected area electron diffraction (SAED). The HRTEM and SAED measurements were performed in a LIBRA 200FE HRTEM, (Carl Zeiss) and the information limit of the HRTEM is 0.13 nm. The surface topography of all the thin films was acquired using Atomic Force Microscope (NT-MDT) in semi-contact mode. Further, the morphology of all the thin film samples were investigated using field emission scanning electron microscope (FESEM, Carl Zeiss Supra 55). The optical characterizations of the thin films were performed in a UV-Vis-NIR spectrophotometer (Hitachi UH4150).

3. **Results and Discussion**

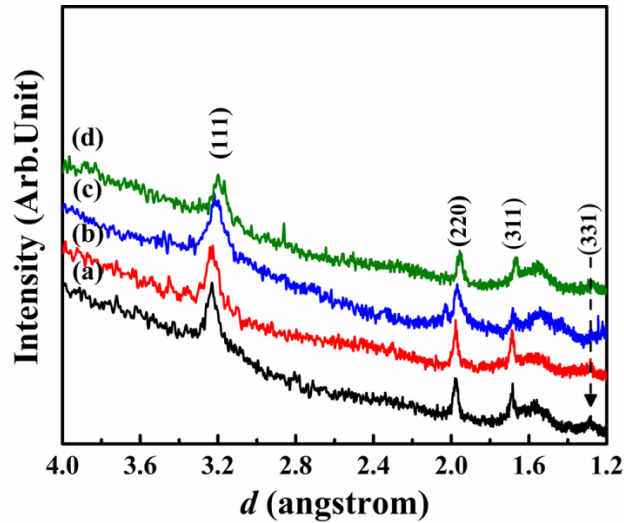

Fig. 1 GIXRD pattern of (a) pristine thin film; and after annealing the pristine thin film at (b) 600 °C; (c) 700 °C and (d) 800 °C for 5 hrs

The GIXRD patterns of the pristine and the annealed thin film samples are shown in Fig. 1. The formation of *poly*-SiGe phase in the pristine thin film can be inferred from the XRD peaks observed at interplanar spacing (*d*) values ~ 3.23, 1.97, 1.68 and 1.28 Å. These peaks correspond to the SiGe (111), SiGe (220), SiGe (311) and SiGe (331) planes, respectively. Moreover, peaks corresponding to *poly*-Ge and *poly*-Si phases are absent in the GIXRD pattern of the pristine thin film. Herein, it may be noted that the crystallization temperature of amorphous bulk Ge ($T_{c(a-Ge)}$) and amorphous bulk Si ($T_{c(a-Si)}$) are ~ 500 °C and 700 °C, respectively [13]. The absence of peaks corresponding to *poly*-Ge shows that all the elemental Ge actively participates in the alloying process. Although peaks corresponding to *poly*-Si are also absent in the pattern, the same cannot be inferred for elemental Si because it is still possible that some fraction of elemental Si continue to remain in amorphous state owing to its higher $T_{c(a\text{-}Si)}$. Consequently, the SiGe thin film that forms is expected to be Ge-rich. Intricate details like the presence of *a*-Si phase can be revealed by an investigation with structural characterization techniques which is both complementary to XRD and is also sensitive to amorphous phases [23]. Further, the XRD peaks corresponding to *poly*-SiGe phase {(111) & (220) planes} shows slight broadening along with a small shift in the peak positions toward lower *d* values, upon annealing the pristine sample at higher $T_a$ (see Fig. 1). The observed broadening is intriguing because when a polycrystalline sample is annealed at higher $T_a$, the crystalline order in the sample generally improves through annealing of various defects and re-crystallization. Disruption of existing long range crystalline order in a sample can occur only in a non-equilibrium process like ion beam irradiation and not in a thermal annealing process. Phase segregation at higher $T_a$ as a factor for the observed broadening can be also ruled out as Si and Ge are completely miscible system. The most probable explanation for the observed broadening in the XRD peaks is nucleation of new additional $Si_{x'}Ge_{1-x'}$ phase with composition different from that of the pristine *poly*-$Si_xGe_{1-x}$ thin film. Moreover, the observed shift in XRD peak positions also indicates a stoichiometric change at higher $T_a$ according to Vegard's law [24]. This inference will be corroborated with other experimental observations and are discussed elaborately later in the article.

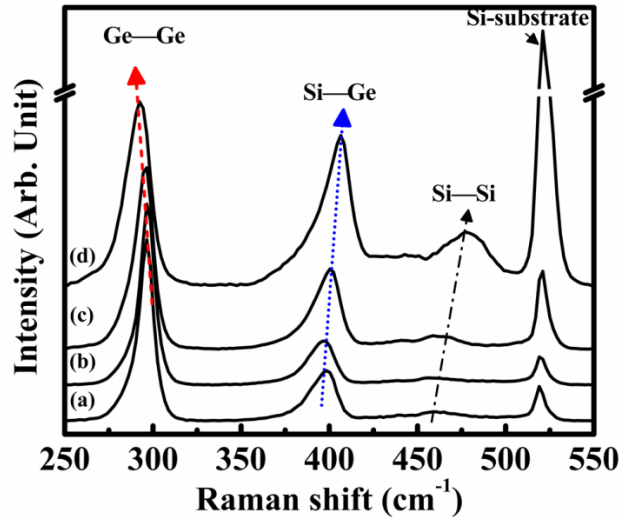

Fig. 2 Raman spectra of the (a) pristine thin film and thin film samples annealed at (b) 600 °C; (c) 700 °C and (d) 800 °C

Raman measurements were performed in the samples to gain further insight into the alloying process. The Raman spectra acquired from the pristine and the annealed thin films are shown in Fig. 2. Apart from the Raman peak at ~ 521 cm$^{-1}$ which arises from the Si substrate [25], the Raman spectra of the pristine thin film exhibit three main peaks centered at ~ 291.6, 398.8 and 459 cm$^{-1}$. These three peaks corresponds to the vibration modes of Ge—Ge, Si—Ge and Si—Si covalent bonds, respectively [26,27]. The peak at ~ 398.8 cm$^{-1}$ confirms the formation of SiGe phase in the pristine thin film. We can also note that the SiGe peak exhibit a small asymmetry in the profile towards the lower energy side. The phonon mode of Si—Ge arises due to the localized vibration of the lighter Si atoms that are bonded to much heavier Ge atoms [28]. Consequently, the peak from Si atoms that are co-ordinated with four Ge atoms (Si-Ge$_4$) appears at lower energy than from Si atoms that are co-ordinated with less than four Ge atoms (i.e., Si-Ge$_3$Si, Si-Ge$_2$Si$_2$, etc). Hence, the observed asymmetry indicates that only a small fraction of Si atoms are co-ordinated with four Ge atoms compared to majority of Si atoms that are co-ordinated with less than four Ge atoms in the pristine thin film. Then, few peaks with low intensities can be also observed in the spatial frequency band between 400 to 500 cm$^{-1}$. These modes arise due to the localized vibration of Si—Si$_{loc}$ bonds in the neighborhoods containing different number of Ge atoms [25,27]. This is a consequence of the complex structure of SiGe, wherein Si and Ge are randomly arranged because of their complete miscibility [23,29]. It may be also noted that the vibrational mode of Si—Si bond due to short range ordering in *a*-Si as well as very small local Si clusters within the SiGe matrix appears in the same frequency band ~ 480 cm$^{-1}$ [26]. In the pristine thin film, the Si—Si mode has its main peak centered at ~ 459 cm$^{-1}$ and have small shoulders at both the lower and higher energies ends. This can be resolved into five peaks at ~ 439.4, 454.8, 463, 471 and 480.3 cm$^{-1}$ as shown in Fig. 3. While the first four peaks are attributed to localized vibration modes of Si—Si, the peak at ~ 480 cm$^{-1}$ predict the presence of *a*-Si phase in the pristine thin film.

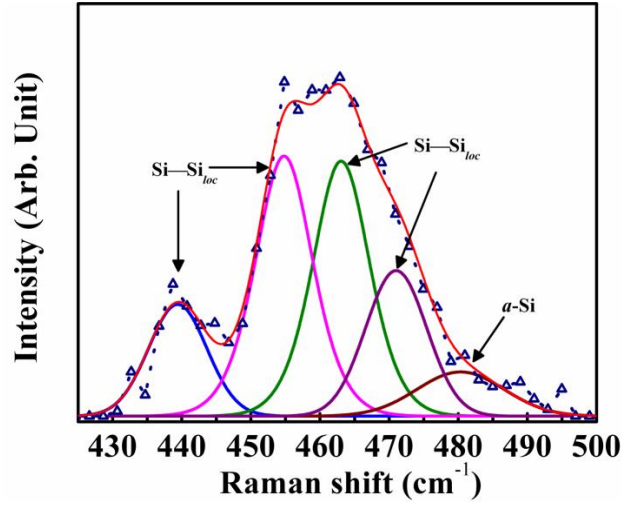

Fig. 3 Raman spectra showing Si—Si mode for the pristine thin film

Further, the Raman peak corresponding to Ge mode exhibits a red-shift towards lower wave numbers as $T_a$ is increased up to 800 °C. Conversely, the Raman peaks belonging to Si—Ge and Si—Si$_{loc}$ modes on the other hand exhibit blue-shifting to higher wave numbers with increase in $T_a$. Red shifts in the Raman peak position of Ge mode with a corresponding blue-shifts in Raman peaks position of SiGe and Si modes are associated with increase of Si content ($x$) in Si$_x$Ge$_{1-x}$ alloys [23,30]. In addition the decrease in intensity ratio of the Ge and SiGe modes ($I_{Ge}/I_{SiGe}$) indicates that the relative Ge content in the SiGe thin film decreases with an increase in $T_a$ [31]. These two observations clearly established that there is a stoichiometric change in the SiGe thin film as a function of $T_a$ in the present study. The stoichiometry of the various samples, estimated from both the Raman peak positions and $I_{Ge}/I_{SiGe}$ ratio are tabulated in Table 1. The calculation of Ge content in the SiGe thin film from the $I_{Ge}/I_{SiGe}$ ratio were performed using the following equation [25,32]

$$\frac{I_{Ge}}{I_{SiGe}} = A \frac{x_{Ge}}{2(1 - x_{Ge})} \qquad (1)$$

where $I_{Ge}$ and $I_{SiGe}$ are the integrated intensity of the Ge—Ge and Si—Ge peaks; $x_{Ge}$ is the Ge content. Herein, the value of '$A$' depends on the laser excitation wavelength and sample details. It has to be calculated for each set of experiments and it can vary from 1 to 3.2 [32]. In the present study, the value of $A = 1.2$ was found by fitting the $x_{Ge}$ values estimated from Raman peak positions using Eqn.1 as shown in Fig. 4.

Table 1: Ge content in the pristine $Si_xGe_{1-x}$ thin film and other samples annealed at various temperature

| Sample details | Ge content ($X_{Ge}$) (from Raman peak positions) | Ge content ($X_{Ge}$) (from $I_{Ge}/I_{SiGe}$ ratio) |
|---|---|---|
| Pristine thin film | 0.86 | 0.79 |
| After annealing at 600 °C for 5 hrs | 0.87 | 0.80 |
| After annealing at 700 °C for 5 hrs | 0.82 | 0.72 |
| After annealing at 800 °C for 5hrs | 0.66 | 0.56 |

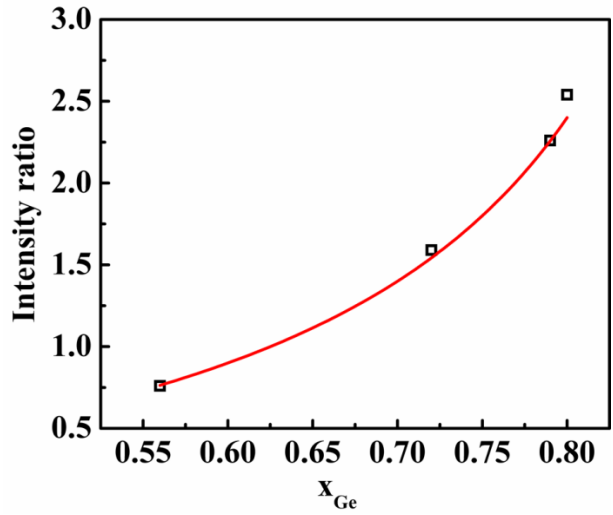

Fig. 4 Plot of intensities ratio of Ge—Ge and Si—Ge modes i.e., $I_{Ge}/I_{SiGe}$ Vs Ge fraction estimated from Raman peak positions, fitted using Eqn. 1

Table 1 shows that the *poly*-$Si_xGe_{1-x}$ alloy thin film is still Ge rich even after annealing at $T_a$ up to 800 °C. However, the Ge to Si ratio has decreased with increase in $T_a$. This increase in relative Si content of the *poly*-$Si_xGe_{1-x}$ alloy thin film (compared to the pristine thin film) with increasing $T_a$ (see Table 1) further strongly supports the existence of *a*-Si along with the SiGe phase in the pristine thin film. The evidence for presence of *a*-Si phases along with the lattice planes of SiGe were observed in the HRTEM images acquired from the pristine thin film shown in Fig. 5(a). Note that the lattice parameter (*a*) and *d* values of SiGe alloy lie between the values of Si and Ge which themselves are closely matched. The SAED pattern of the pristine thin film is shown in Fig. 5(b). The pattern shows a series of concentric diffraction rings over a diffuse halo background. While the concentric diffraction rings correspond to *poly*-SiGe phase, the diffuse diffraction ring in the background indicates the presence of *a*-Si phase. (see Fig. 5(b)). These *a*-Si phase acts as reservoir from which Si atoms diffuse into the SiGe at higher $T_a$ and increase the Si content of the alloy thin film. Subsequently, when Si atoms from the *a*-Si phase diffuse into SiGe at higher $T_a$( > 500 °C), the weak Raman peak due to Si—Si mode in *a*-Si slowly disappears while the Si—Si$_{loc}$ mode blue shifts to higher frequencies along with an increase in intensity (see Fig. 2). The diffusion of Si into SiGe continue until all the *a*-Si phase finally gets completely dissolve into SiGe phase at $T_a$ = 800 °C. The HRTEM images obtained from the film annealed at $T_a$ = 800 °C are shown in Fig. 5(c). It only shows crystalline planes belonging to $d_{(111)}$ and $d_{(220)}$ planes of SiGe. The corresponding SAED pattern of the film annealed at $T_a$ = 800 °C is shown in Fig. 5(d). The diffuse halo diffraction ring observed in the pristine thin film due to the presence of *a*-Si has disappeared (Fig. 5(a)). Instead, well defined diffraction ring pattern with discrete diffraction spots corresponding to *d* values of *poly*-SiGe phase are observed.

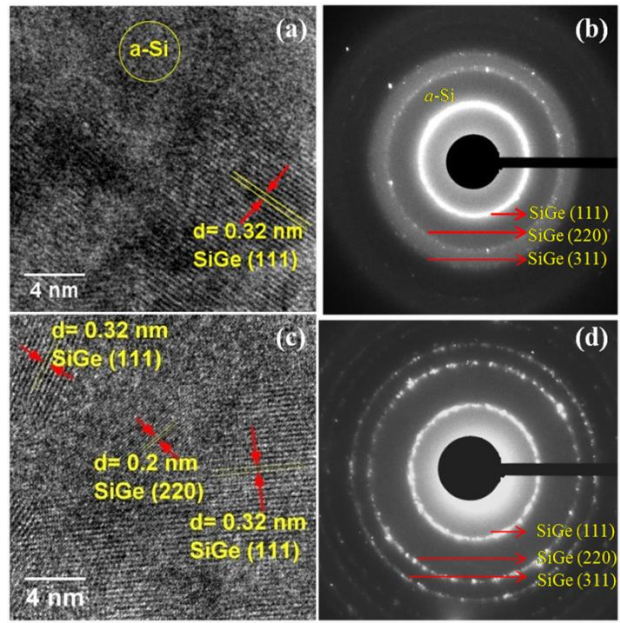

Fig. 5 HRTEM imaging and SAED pattern: (a) & (b) pristine thin film; and (c) & (d) sample annealed at 800 °C for 5 hrs

The optical properties of these *poly*-SiGe alloy thin films were then investigated using a UV-Vis-NIR spectrophotometer. The energy band gaps of these alloy thin films were calculated using Tauc's formalism [33]. The Tauc plots of these thin films are shown in Fig. 6 and the values estimated from these plots are tabulated in the Table 2. The pristine thin film has energy bandgap $E_{g(SiGe)}$ ~ 0.75 eV which is close to and slightly higher than $E_{g(Ge)}$ ~ 0.67 eV. This is in consistence with our results which shows that the pristine SiGe alloy thin film is Ge rich. Then, as the Si fraction ($x$) in the $Si_xGe_{1-x}$ alloy thin film increases with increase in $T_a$, $E_{g(SiGe)}$ also increases and is shifted toward $E_{g(Si)}$. However, at $T_a \geq 600$ °C, two distinct linear regions could be observed in the Tauc plots (see Fig. 6). These two linear regions signifies onset of two different absorption processes. The origin of two absorption edges in the films annealed at $T_a \geq 600$ °C can be understood on the basis of the microstructure of the SiGe alloy thin film and the diffusion characteristics of Si upon annealing.

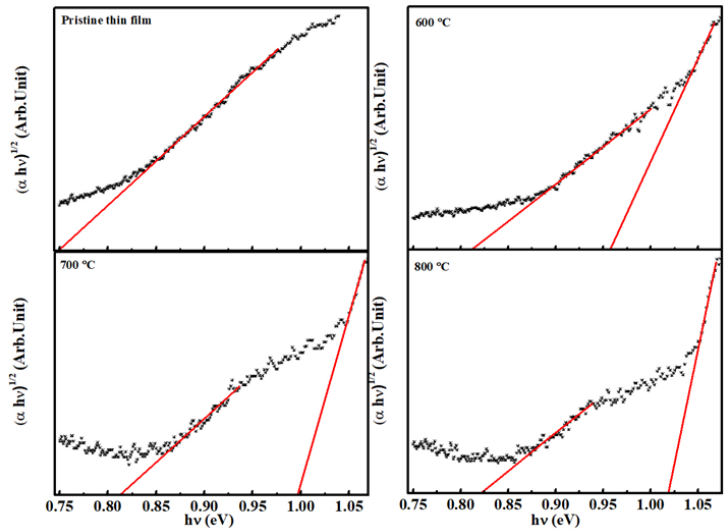

Fig. 6 Tauc plot of the pristine of thin film and the samples annealed at various temperatures

Table 2: Energy band gap and *rms* roughness values for the pristine thin film and other samples annealed at various temperatures

| Sample details | Band gap (eV) | Rms roughness (nm) (scan area: 2 x 2 µm2) |
|---|---|---|
| Pristine thin film | 0.75 | 5.49 |
| After annealing at 600 °C for 5 hrs | 0.81, 0.96 | 3.76 |
| After annealing at 700 °C for 5 hrs | 0.82, 0.99 | 2.52 |
| After annealing at 800 °C for 5 hrs | 0.83, 1.02 | 2.47 |

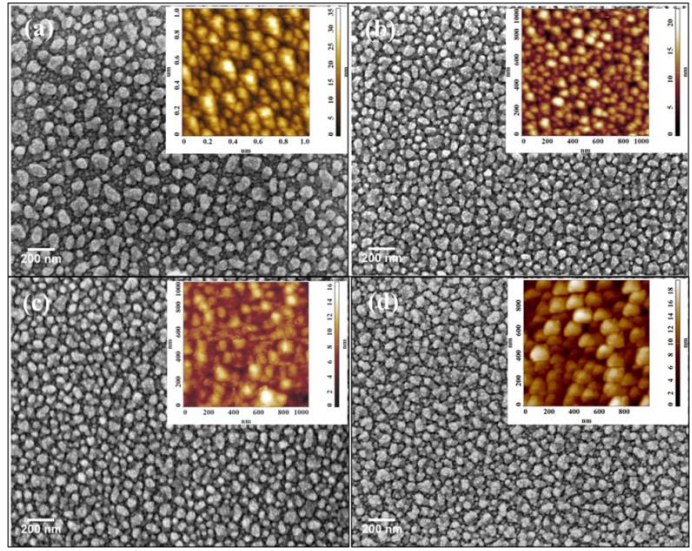

Fig. 7 FESEM micrograph of (a) Pristine thin film; and after 5 hrs annealing at (b) 600 °C; (c) 700 °C and (d) 800 °C. AFM micrograph showing corresponding surface topographies are shown as insets

The effect of the crystallization annealing on the microstructures of the SiGe alloy thin film were investigated with FESEM and AFM. The SEM micrographs displaying the surface morphologies of the pristine thin film and films annealed at various temperatures are shown in Fig. 7. The SEM micrographs show that the surface morphologies of all the films consist of fine grainy microstructural features with varying surface topography. AFM micrographs displaying the scanned surface topography of the corresponding film surfaces are shown as inset for each sample in Fig. 7. The surface topography of the pristine thin film is characterized by a fine grain growth over which bigger agglomerates of finer grains are uniformly distributed. The average length of these agglomerates measure ~ 70 nm. This topography is a consequence of the growth process. During the sequential evaporation at $T_s = 500$ °C, while the Ge sub-layer crystallizes into *poly*-Ge as $T_{c(a-Ge)}$ is below the $T_s$, the Si sub layer does not as $T_{c(a-Si)}$ (≈ 700 °C) is considerably higher than the $T_s$. Herein, the diffusion co-efficient of Si in Ge ($D_{Si \rightarrow Ge}$) is much higher compared to that of Ge in Si ($D_{Ge \rightarrow Si}$) in Si-Ge system [34]. This asymmetry in diffusion exists because Si atoms being smaller can diffuse into Ge lattice with ease compared to diffusion of much larger Ge in Si lattice. Now at $T_s = 500$ °C, the lattice contribution to diffusion is very small ($D_{Si \rightarrow Ge}$ below $10^{-19}$ cm$^2$s$^{-1}$) and primary contribution to diffusion stems from grain boundaries (GB) and interface diffusions [35]. As such, the dissolution of the *a*-Si layer sandwiched between two *poly*-Ge layers that initiate the alloying of Si and Ge proceeds primarily via GB diffusion of Si into the Ge layers. For the topmost *a*-Si layer, diffusion of Si→Ge layer (which has effectively become SiGe now) through GB mechanism continues to dissolve *a*-Si. However, since magnitude of surface diffusion is higher than GB diffusion, surface diffusion emerge as a competing process and it became the dominant diffusion mechanism for the incoming Si ad-atoms at the film's surface [34]. In such scenario, the incoming Si ad-atoms can diffuse along the surface due to high surface diffusivities and forms agglomerates on the film's surface (see SEM micrograph, Fig. 7). The surface of the pristine thin film has a rough surface topography due to the presence of these agglomerates (see AFM micrograph, Fig. 7). The surface topography further evolves when the thin films are subjected to post deposition annealing in $N_2$ ambient at $T_a$ ranging from 600 to 800 °C. The *rms* surface roughnesses value for various thin films, estimated from the AFM micrographs are tabulated in Table 2.

When the pristine thin film is annealed at $T_a > 500$ °C, the *a*-Si in the film is continuously subsumed into SiGe phase through diffusion of Si into SiGe. This process leads to smoothening of the surface topography which is largely due to Si agglomerates on the surfaces. Subsequently, the *rms* surface roughnesses decreases with increasing $T_a$ (Table 2). The quantitative understanding of this diffusion process is difficult due to the sparse data that is available on GB diffusion of Si (&Ge) →*poly*-Ge(& Si) as well as Si (&Ge)→*poly*-SiGe. Nevertheless, important inference can be drawn from the data available for diffusion of Si and Ge in epitaxial Si, Ge and their alloys. Herein, the lattice or bulk diffusion of Si→Ge becomes significant only at high temperature near $T_{c(a\text{-Ge})}$. For Ge→Si, it become significant at much higher temperature near $T_{c(a\text{-Si})}$. At $T_a = 700$ °C, the bulk diffusion co-efficient of Si→Ge ($D_{Si\to Ge}$) ≈ $10^{-16}$ cm$^2$s$^{-1}$ and it becomes ≈ $10^{-14}$ cm$^2$s$^{-1}$ at $T_a = 800$ °C [35]. The value for Ge→Si is very small with $D_{Ge\to Si}$ ≈ $10^{-22}$ cm$^2$s$^{-1}$ at $T_a = 700$ °C and becomes ≈ $10^{-14}$ cm$^2$s$^{-1}$ only at $T_a > 1100$ °C [36]. Compare to $D_{Si\to Ge}$ or $D_{Ge\to Si}$, magnitude of diffusion co-efficient for (Si & Ge)→SiGe ($D_{(Si\&Ge)\to SiGe}$) are several orders lower [35,37]. Moreover, $D_{Si\to SiGe}$ also reduces when Si fraction (*x*) in the Si$_x$Ge$_{1-x}$ alloy increases [38]. $D_{Si\to SiGe}$ decrease from ~$10^{-17}$ to ~ $10^{-18}$ cm$^2$s$^{-1}$ at $T_a = 800$ °C when *x* increases from 0.45 to 0.70 in Si$_x$Ge$_{1-x}$ alloy, respectively [37]. This indicates that as the Si content (*x*) in the Si$_x$Ge$_{1-x}$ thin film increases due to continued diffusion of Si into *poly*-SiGe during post annealing at $T_a > 500$ °C, subsequent diffusion of additional Si atoms is retarded. Among other factors, this retardation is due to the decreasing concentration gradient of Si which is the main driving force for diffusion.

At $T_s$ or equivalently $T_a$ = 500 °C, the competing diffusion processes that transport Si atoms from the *a*-Si phase on the surface into the growing *poly*-SiGe thin film phase are lattice and GB diffusions. Only at much higher $T_a$, contribution from lattice gains more importance than the GB contribution. This is particularly true in cases where GB densities decrease due to rapid grain growth following recrystallization of the material at high $T_a$. In the present study, there are no indication of any rapid or significant grain growth up to $T_a$ = 800 °C. In such scenario, it is reasonable to expect GB diffusion to remain as a dominant diffusion mechanism up to the $T_a$ range used in this study (800 °C), in addition to the lattice diffusion which increases with $T_a$. Also note that diffusion rate through GB are faster than through lattice [34]. Hence, the diffusion of Si into SiGe thin film upon annealing at $T_a$ > 500 °C in the present study is a complex process because of the significance of the GB diffusion. GB diffusion being faster transport and homogenize the Si concentration in the SiGe thin film, thereby increasing the overall Si content of the *poly*-$Si_xGe_{1-x}$ phase compared to the pristine thin film. The energy bandgap appearing at lower energy in the range ~ 0.81 - 0.83 eV in the optical measurement corresponds to this phase. On the other hand because of its slower $D_{Si \to SiGe}$, the lattice diffusion increases the Si content of the SiGe thin film only near the surface, thereby introducing a concentration gradient. The energy bandgap observed at higher energy range ~ 0.96 – 1.02 eV corresponds to this phase. The net result is the evolution of *poly*-$Si_xGe_{1-x}$ alloy thin film having a Si richer *poly*-$Si_{x'}Ge_{1-x'}$ phase near the surface ($x' > x$). The diffusion process in the present film is illustrated in the schematic shown in Fig. 8.

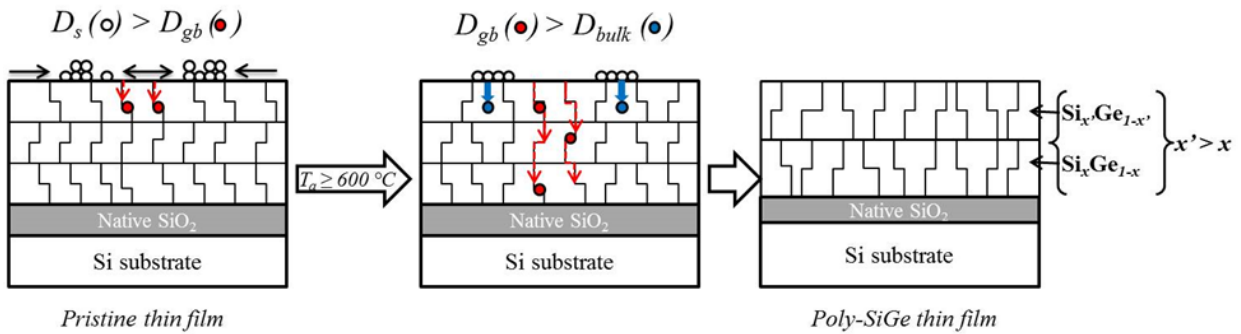

Fig. 8 Schematic illustration of the diffusion process of Si when the pristine thin film is annealed at $T_a \geq$ 600 °C up to 800 °C; $D_s$, $D_{gb}$ and $D_{bulk}$ are the surface, grain boundary and bulk diffusivities

These Ge rich *poly*-SiGe thin films have the potential to increase the absorption efficiency particularly in the longer wavelength i.e., IR region of the solar spectrum. *Poly*-SiGe thin film were also successfully deposited on Corning glass substrate at $T_s$ = 500 °C using e-beam evaporation. The GIXRD pattern and Raman spectrum showing the *poly*-SiGe phase in the sample deposited on glass is shown in Fig. 9. This e-beam deposited *poly*-SiGe thin film using optimized in-situ heat treatment of the substrate is compatible with low temperature fabrication process for low cost applications.

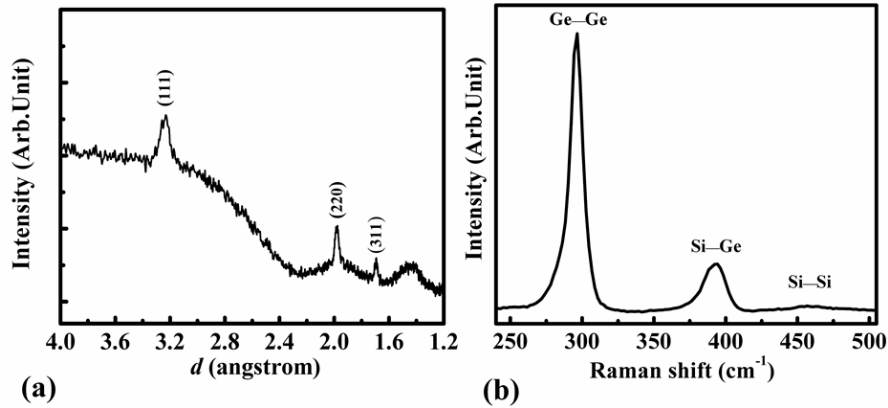

Fig. 9 GIXRD pattern and Raman spectra of the pristine thin film deposited on Corning glass substrate

### 4. Conclusion

Ge rich *poly*-SiGe alloy thin films were obtained on Si and Corning glass substrates at low temperature of 500 °C which is near the crystallization temperature of Ge. This was achieved by *in-situ* heating of the substrates at 500 °C during e-beam evaporation of Ge and Si. The effect of post deposition annealing on the stoichiometry and energy band gap were investigated. Two absorption edges could be observed in the alloy thin film after annealing at temperatures ≥ 600 °C. The origin of these edges was attributed to the differential diffusion rates of Si atoms through the grain boundaries and the lattice of the *poly*-SiGe thin films. These Ge rich *poly*-SiGe thin films can potentially increase the solar absorption efficiency in the IR region owing to their lower band gap.

**Acknowledgements**

The authors would like to acknowledge Dr. A. K Sinha (*RRCAT, Indore*), Dr. Neha Sharma and Dr. R. Pandian for their help during synchrotron experiments, UV-Vis-NIR and SEM measurements, respectively. The authors would also like to thank Directors, MSG and IGCAR for supporting this work.